# Single-Exciton Gain and Stimulated Emission across the Infrared Telecom Band from Robust Heavily-doped PbS Colloidal Quantum Dots

Sotirios Christodoulou*[1,†], Iñigo Ramiro[1, †] , Andreas Othonos[2], Alberto Figueroba[1], Mariona Dalmases[1], Onur Özdemir[1], Santanu Pradhan[1], Grigorios Itskos[3], Gerasimos Konstantatos*[1,4]

1. ICFO-Institut de Ciencies Fotoniques, The Barcelona Institute of Science and Technology, 08860 Castelldefels (Barcelona), Spain
2. Laboratory of Ultrafast Science, Department of Physics, University of Cyprus, Nicosia 1678, Cyprus
3. Experimental Condensed Matter Physics Laboratory, Department of Physics, University of Cyprus, Nicosia 1678, Cyprus
4. ICREA – Institució Catalana de Recerça i Estudis Avançats, Lluis Companys 23, 08010 Barcelona, Spain

[†] these authors contributed equally in the work

Gerasimos.konstantatos@icfo.es , Sotirios.chrsitodoulou@icfo.es

## Abstract

**Materials with optical gain in the infrared are of paramount importance for optical communications, medical diagnostics and silicon photonics. The current technology is based either on costly III-V semiconductors that are not monolithic to silicon CMOS technology or Er-doped fiber technology that does not make use of the full fiber transparency window. Colloidal quantum dots (CQD) offer a unique opportunity as an optical gain medium in view of their tunable bandgap, solution processability and CMOS compatibility. The 8-fold degeneracy of infrared CQDs based on Pb-chalcogenides has hindered the demonstration of low-threshold optical gain and lasing, at room temperature. We demonstrate room-temperature, infrared, size-tunable, band-edge stimulated emission with linewidth of ~14 meV. Leveraging robust electronic doping and charge-exciton interactions in PbS CQD thin films, we reach gain threshold at the single exciton regime representing a four-fold reduction from the theoretical limit of an eight-fold degenerate system, with a net modal gain in excess of 100 cm$^{-1}$.**

The potential for narrower linewidths[1] and the lower-than-bulk degeneracy[2] has led to dramatic progress towards successful demonstration of optical gain[3], stimulated emission[4] and lasing[5,6,7] in the visible part of spectrum utilizing CdSe-based CQDs. Therefore, low threshold, band-edge amplified spontaneous emission (ASE) in CQDs has been at the center of intensive research over recent years as a prerequisite towards the demonstration of CQD lasing.[7,8,9] Engineered CQDs with suppressed Auger[4] and photodoping[10] have allowed the realization of low threshold ASE and lasing[11] at the single exciton regime, in the visible, for CdSe CQD systems that possess a two-fold degeneracy value.[12,13] Nevertheless, low-threshold band-edge ASE in the near-infrared (NIR), based on colloidal quantum dots, has remained a challenge due to the high degeneracy of Pb-chalcogenide CQDs. PbS CQDs is the material of choice for solution processed infrared optoelectronics with successful demonstrations in LEDs,[14] solar cells[15,16] and photodetectors[17,18]. The 8-fold degeneracy of PbS(e) CQDs, however, has hindered the demonstration of low-threshold optical gain and ASE, at room temperature, in the infrared across the telecommunications wavelength band. [19,20]

We posited that a CQD film, robustly n-doped in the heavy doping regime, can address this challenge by utilizing the doping electrons present in the first excited state of the CQDs (conduction band) to reach the population inversion condition at reduced pumping fluence. Previously the use of charged CQDs has resulted in lowering the ASE threshold in CdSe-based systems employing photodoping,[10,11] an approach that requires inert conditions and that is reversible in time, preventing thus the demonstration of robust, permanently doped CQDs and devices thereof. Therefore, we sought to develop a robust electronic n-doping method for PbS CQDs in the heavy doping regime based on atomic substitutional doping. Over the years, various doping strategies have been employed to CQD to tune their electronic properties.[21] Prior reports of heavily doped PbS(e) QDs have relied on aliovalent cation substitution: $Ag^+$ substitution of $Pb^{2+}$ induces p-type doped PbSe[22] CQDs, while the substitution of $Pb^{2+}$ by $In^{3+}$ leads to n-doped PbSe[23] CQDs. Remote transfer of electrons from cobaltocene molecules as well as photochemical doping have also been reported as doping mechanisms for n-type PbSe CQDs[24,25]. However, none of the above approaches have led to robust permanent doping, preventing thus their use in devices. We instead posited that aliovalent anionic substitution may lead to a more robust doping scheme. DFT calculations have theoretically predicted n-type doping upon iodide substitution, yet not experimentally observed[26]. Here we sought to use PbS (100) surfaces as a platform to facilitate heavy doping through iodide substitution (Figure 1a). Our selection has been guided by the DFT calculations shown in Figure 1b taking into account also the exposure of such facets on the surface of large PbS CQDs to enable the implementation of the doping. Iodide binding on (111) surfaces, on the other hand, serves as a passivant without causing any strong doping effects[27] (Supplementary Section 1).

Bleach of the first exciton transition as a result of Pauli blocking is one of the main signatures of successful population of the CB.[28,29] However, bare iodide-exchanged samples did not show any absorption bleaching as the DFT calculations predicted (see Supplementary Section 2.1). The absence of heavy doping, evidenced by the lack of bleaching in the absorption measurements is ascribed to the presence of oxygen and water in the film upon exposure to ambient conditions. Oxygen and water are efficient oxidants and have been reported as effective p-type dopants in lead chalcogenides.[30,31] In order to preserve heavy n-doping in our films under ambient conditions, we submitted our samples to atomic layer deposition (ALD) of alumina with a two-fold purpose: impeding oxygen to further incorporate in the film, and preventing oxidation caused by oxygen/water adsorbates pre-existent in the films upon their formation. Infilling with alumina (supported by XPS results in Supplementary Section 3), hence, is crucial to preserve robust n-type doping provided by iodide ligand-exchange procedure. We have observed that only the samples that have undergone iodide doping and ALD encapsulation demonstrate optical bleaching, herein in this study are assigned as doped

PbS CQD and the samples only with iodine substitution as undoped samples. Control samples of thiol-based ligand exchange chemistry did not show any signatures of heavy doping upon ALD encapsulation, further corroborating the role of iodine as the n-type dopant (see Supplementary section 2.1). Absorption spectra at shorter wavelengths show that the higher energy states of the CQDs remain unaffected by the doping process (see Supplementary section 2.2)

Figure 1d shows exciton peak bleach after alumina deposition, indicating that the ALD process was successful in maintaining heavy doping at ambient conditions. We have been able to quantify the doping (number of electrons in the CB per dot) in our samples optically, by analyzing the bleach in the absorption (see Methods), while we further confirm the doping level electrically via field effect transistor measurements (see Supplementary Section 4), as shown in Figure 1e. Small dots have an octahedral shape with Pb-rich (111) facets, while, as the dot diameter increases, their morphology evolves progressively to a cuboctahedron that has six sulphur-rich (100) facets.[32] Because (100) surfaces are progressively exposed with increasing PbS CQD size, the doping efficacy of the process increases with the size of the dots. Figure 1e shows that particles smaller than 4 nm in diameter do not undergo this doping process due to the lack of (100) exposed facets, whereas in particles with an exciton peak of more than 1800 nm (~7.5 nm in diameter) the conduction band is fully filled with 8 electrons (Figure 1e) in accordance with the eight-fold degeneracy of PbS CQDs. This is also evidenced by ultraviolet photoelectron spectroscopy (UPS) (see Supplementary section 5) according to which the Fermi level of small high-bandgap CQDs upon doping does not reach close to the conduction band (CB) in contrast to the large CQDs where the Fermi level crosses the CB, suggestive of heavily doped material. It is noteworthy that the doped CQD films are stable at room temperature and ambient conditions for a period of more than 2 months (see Supplementary Section 6). To shed further evidence on the role of iodine substitution in the doping process we have performed

XPS measurements of PbS CDQ films, both with the original oleate ligands and with exchanged iodide ligands (see Supplementary Section S7). Quantitative analysis of the lead and sulphur data (see Supplementary Table S2) show that the Pb/S ratio increases after ligand exchange, consistent with substitution of sulphur by iodine. Moreover, as the particle size increases (wherefore more sulphur atoms are available at the surface) the relative increase in the Pb/S ratio after ligand exchange is more pronounced. In contrast, this doping mechanism - via iodide substitution- becomes ineffective in small PbS CQDs whose surface comprises Pb-rich (111) exposed facets [33,34].

To verify our hypothesis of reaching single exciton gain threshold in doped PbS CQD films we performed transient absorption (TA) studies in undoped PbS CQD films as well as a series of doped PbS CQD films with variable initial electron occupancy doping $\langle N \rangle_D$, determined by their size (Figure 1e). Henceforth, we have employed a hybrid ligand exchange treatment based on $ZnI_2$/MPA that caters for equally effective doping (as shown in Figure 1e) as well as serves better passivation of the CQDs and higher photoluminescence, as previously reported.[14] The undoped PbS CQD films demonstrate optical gain threshold $\langle N \rangle_{thr}$ -expressed in excitons per dot of four- as expected from their 8-fold degeneracy (Figure 2a). Upon doping, the $\langle N \rangle_{thr}$ reduces and for the case of initial doping $\langle N \rangle_D$ of 5.4 the $\langle N \rangle_{thr}$ reaches a value of 0.9 excitons per dot (Figure 2b). This four-fold reduction of the gain threshold upon doping outperforms the two-fold reduction reported in CdSe based CQD systems[14]. By varying the initial doping of the CQD films, according to the size-doping dependence shown in Figure 1c, we have measured $\langle N \rangle_{thr}$ values of 4.3, 2.1, 1 and 0.9 excitons per dot for QD sizes with diameter (initial doping) of: 5.0 nm ($\langle N \rangle_D = 1.4$), 5.6 nm ($\langle N \rangle_D = 3.4$), 5.9 nm ($\langle N \rangle_D = 4.4$) and 6.2 nm ($\langle N \rangle_D = 5.4$) respectively (Figure 2c). This finding further corroborates our hypothesis of the effect of doping on the optical gain threshold. It is noteworthy that while the gain threshold measured in the undoped films is in good agreement with the value predicted by simple theoretical and Poisson-statistical models (see Supplementary section S8) the experimentally determined gain threshold for the doped QDs is distinctly lower that the one predicted by those models. However it comes in good agreement when taking into account charged–exciton and exciton-exciton interactions[15], which lower the gain threshold in heavily doped PbS CQDs (see Supplementary section S8).

The corresponding transient absorption traces of both the undoped and doped films (Figure 2d-e) have been collected at the gain wavelength. We notice that already below the gain threshold the undoped sample shows multi-exponential dependence while above gain both doped and undoped samples are characterized by a recombination channel that saturates with $-\Delta\alpha/\alpha_{PbS}$ and whose lifetime component is excitation independent. In order to elucidate the multi-exciton and charge exciton dynamics we used as a case study ~5.7 nm PbS QDs and we collected the TAS for the undoped and doped ($\langle N \rangle_D = 3.6$) films probing the relaxation dynamics up to 1 ns (shown in Supplementary section S9). In our model we have considered that the photogenerated exciton occupancy distribution follow Poisson statistics. By fitting the lifetime traces of the undoped samples at low exciton occupancy ( $\langle N \rangle < 1$ ) we extract the single exciton lifetime on the order of 1.5 ns as well as the Auger lifetime that follows statistical scaling with a measured biexciton lifetime of ~200 ps. Therefore, below the gain threshold the relaxation mechanism is due to Auger recombination as well as singe exciton relaxation.[35] Above the gain threshold there is a rise of an excitation independent component with a lifetime of ~~~35~~25-30 ps for large CQDs (6.2 nm) and 15 ps for smaller sized dots (5.0 nm) that saturates following the $-\Delta\alpha/\alpha_{PbS}$ and which we attribute as the gain lifetime as it does not follow the Auger scaling. It is noteworthy that in the gain regime the TAS dynamics do not reproduce the Auger scaling dependence (Fig. S9.1). We instead measure a strong relaxation component of 25-35 ps (assigned as the gain lifetime) despite the fact that at those conditions (excitation of 4 excitons per dot) Auger scaling, taking into account Poisson statistics, predicts much faster

Auger lifetime components for the population of dots that are excited with more than 4 excitons/dot (see table in SI section 9). This suggests that the stimulated emission channel is favorable over non-radiative Auger recombination for carriers when in the gain regime. The mechanism behind this intriguing observation is still unclear and merits further investigation.

In the doped CQD films, at fluences below the gain regime, Auger recombination takes place even at the single exciton occupancy excitation as a result of the dopant electrons at the conduction band. Our experimental results however point out the existence of two recombination channels, one following the Auger scaling and a longer one that we attribute to single charged-exciton recombination (see Supplementary section S9). TAS analysis in the gain regime (Supplementary section S9, Fig. S9.2) discloses the presence of an excitation-independent recombination pathway that does not follow the Auger scaling, similar to the undoped films, and that we assign as the gain lifetime of the system. To further elucidate the dynamics of the system we have both calculated and extracted experimentally the Auger lifetime for a series of doped CQDs at their corresponding gain threshold. Figure 2f plots the experimentally measured excitation independent component assigned as the gain lifetime along with the experimentally extracted Auger lifetime at gain threshold. The experimentally extracted Auger lifetime matches well with the calculated Auger lifetime, when we consider only the nominal exciton occupancy of the dots at their corresponding thresholds (i.e. when ignoring the faster Auger components from the dots with larger exciton occupancy due to Poisson statistics). Interestingly, we note that doped QDs at threshold possess longer Auger lifetimes than gain lifetimes thanks to reaching the gain regime at lower exciton occupancies. The Auger lifetime on threshold is found to increase for a doping occupancy up to 3 electrons per dot. Beyond this point Auger lifetime accelerates due to increasing number of free carriers in the dots. Through-out the experimental conditions of this work the gain lifetimes remain in the range of 20 – 30 ps thus faster compared to the experimentally determined Auger lifetimes, which supports that the stimulated emission is the main de-excitation channel.

Optical gain is a prerequisite for stimulated emission. Having achieved this, next we performed amplified spontaneous emission (ASE) measurements of our thin films. In line with the TA measurements we observed stimulating emission from both doped and undoped samples (see Supplementary Section 11). We calculated the stimulated emission occupancy threshold $<N>_{thr}$ by fitting the integrated PL spectra (see Supplementary Section 12) from the power dependent S-curves of all the samples. In Figure 3 a,b we plot the integrated ASE peak as a function of exciton occupancy and the respective power dependence measurement of two representative PbS sizes of 5.4nm ($<N>_D = 2.7$) and 5.8nm ($<N>_D = 4.0$) (Figure c-f). All the undoped samples have an $<N>_{thr}$ of 4, in agreement with the TAS measurements of Figure 2, and the ASE signal is saturated when 8 electrons have fully populated the conduction band. Power dependence measurement of those samples are shown in Figure c-f positioning the stimulating emission peak in wavelengths above 1500 nm. The sharp ASE peak has an average FWHM of 14 meV, characteristic of a stimulated emission process and comparable with reported values in the visible from CdSe-based systems.[10] To our knowledge, this is the first report of CQD ASE in the infrared characterized by ASE saturation and narrow linewidth, essential features of ASE that had remained elusive hitherto[19,36] . The stimulated emission threshold occupancy is summarized in Figure 3g. The undoped samples preserve a constant value of $<N>_{thr}$ of 4 independent of their size, whereas in the case of doped CQDs increasing their size -and thereby the initial doping occupancy- the stimulated emission threshold decreases to a minimum value of 1.3 excitons in good agreement with the transient absorption measurements.

A figure of merit of paramount importance for applications in optical amplification and lasing is the net modal gain of the material. We have experimentally measured the net modal gain $g_{modal}$ using the variable stripe length (VSL) technique (see Supplementary Section 13). We report an average $g_{modal}$ of 30 $cm^{-1}$ nearly constant for all the undoped samples (Figure 3h). Upon doping, $g_{modal}$ increases up to a value of 114 $cm^{-1}$ for an $\langle N \rangle_D$ of 4, when the conduction band is half filled. This modal gain value outperforms Er-doped fiber systems with values in the range of 0.01 – 0.1 $cm^{-1}$ and compares favourably to costly epitaxial III-V multi quantum well and quantum dot systems[37]. We demonstrate tunable coverage across the optical fiber communication spectrum (Figure 3i) from a solution processed material, extending beyond the spectral coverage of Er-doped fiber systems. The tunable spectral coverage across the infrared taken together with the high gain values and the low threshold represent a significant advance towards the development of infrared CQD solution processed lasers. Further improvements in doping efficacy may lead in the future towards threshold-less lasing in such doped CQD films. The latter taken together with future developments of engineered lead chalcogenide CQDs with suppressed Auger[16] and the appropriate selection of thermal dissipation schemes[9] can pave the way towards CW lasing (predictive calculations on this are provided in Supplementary section 14).

**Acknowledgements**

The authors acknowledge financial support from the European Research Council (ERC) under the European Union's Horizon 2020 research and innovation programme (grant agreement no. 725165), the Spanish Ministry of Economy and Competitiveness (MINECO) and the 'Fondo Europeo de Desarrollo Regional' (FEDER) through grant TEC2017-88655-R. The authors also acknowledge financial support from Fundacio Privada Cellex, the program CERCA and from the Spanish Ministry of Economy and Competitiveness through the 'Severo Ochoa' Programme for Centres of Excellence in R&D (SEV-2015-0522). S.C. acknowledges support from a Marie Curie Standard European Fellowship (NAROBAND, H2020-MSCA-IF-2016-750600). I. Ramiro acknowledges support from the Ministerio de Economía, Industria y

Competitividad of Spain via a Juan de la Cierva fellowship. We are also thankful to Dr. Kaifen Wu for fruitful discussions.

FIGURES

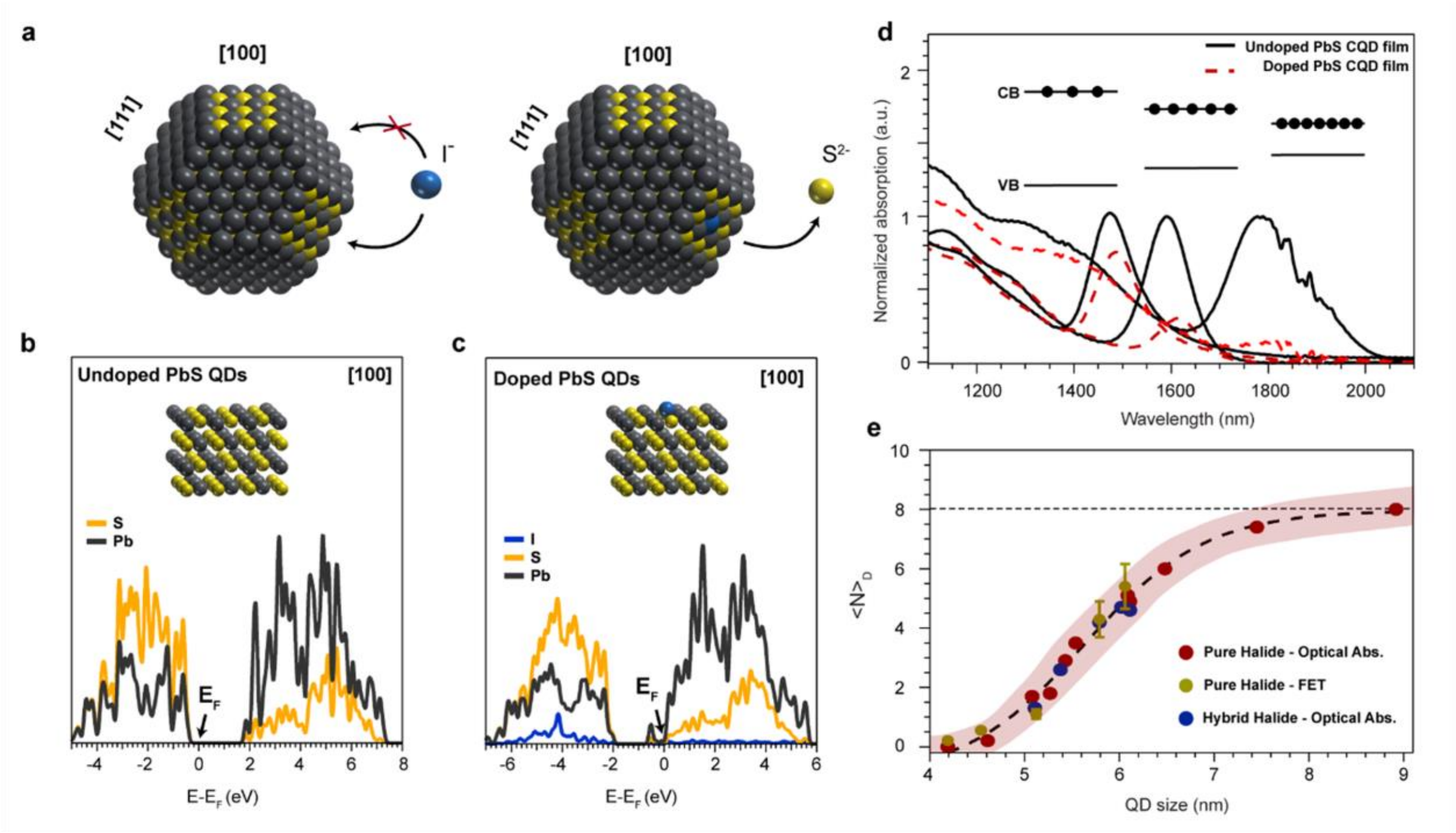

**Figure 1. a,** Schematic representation of the $S^{2-}$ substitution to $I^-$ in (100) surface in large, cuboctaehedral-shaped PbS CQDs **b,c** Calculated density of states (DOS) of the (100) surface before and after $I^-$ substitution showing that the Fermi level, $E_F$ is shifted to the conduction band **d**, Absorption spectra of two representative PbS CQD films, before and after doping. The CQD sizes are 5.5 nm and 6.1 nm with respective exciton peaks at1480 nm and 1580 nm (solid black lines). After doping of the CQD films the absorption bleaches (red dash lines) **e**, The number of electrons in the conduction band, upon doping, depends on the size of the CQD due to the degree of (100) surface presence that enables doping. Red dots represent the experimentally extracted number of electrons from measuring the bleaching of the absorption of the films at the exciton peak with the use of 1-ethyl-3-methylimidazolium iodide (EMII) for ligand exchange while blue dots represent the $ZnI_2$/MPA hybrid ligand treated films that are used for ASE and gain measurements. Yellow dots indicate the extracted doping values from electrical FET measurements of transistors based on EMII-treated CQDs. Both ligand treatments are equally effective in the doping of the CQDs. The red area represents the doping range variation based on size distribution of the CQDs indicating even the doping distribution across the films is ± 0.5 electrons. The data have been fitted with a sigmoidal function (black dash line) as a guide to the eye.

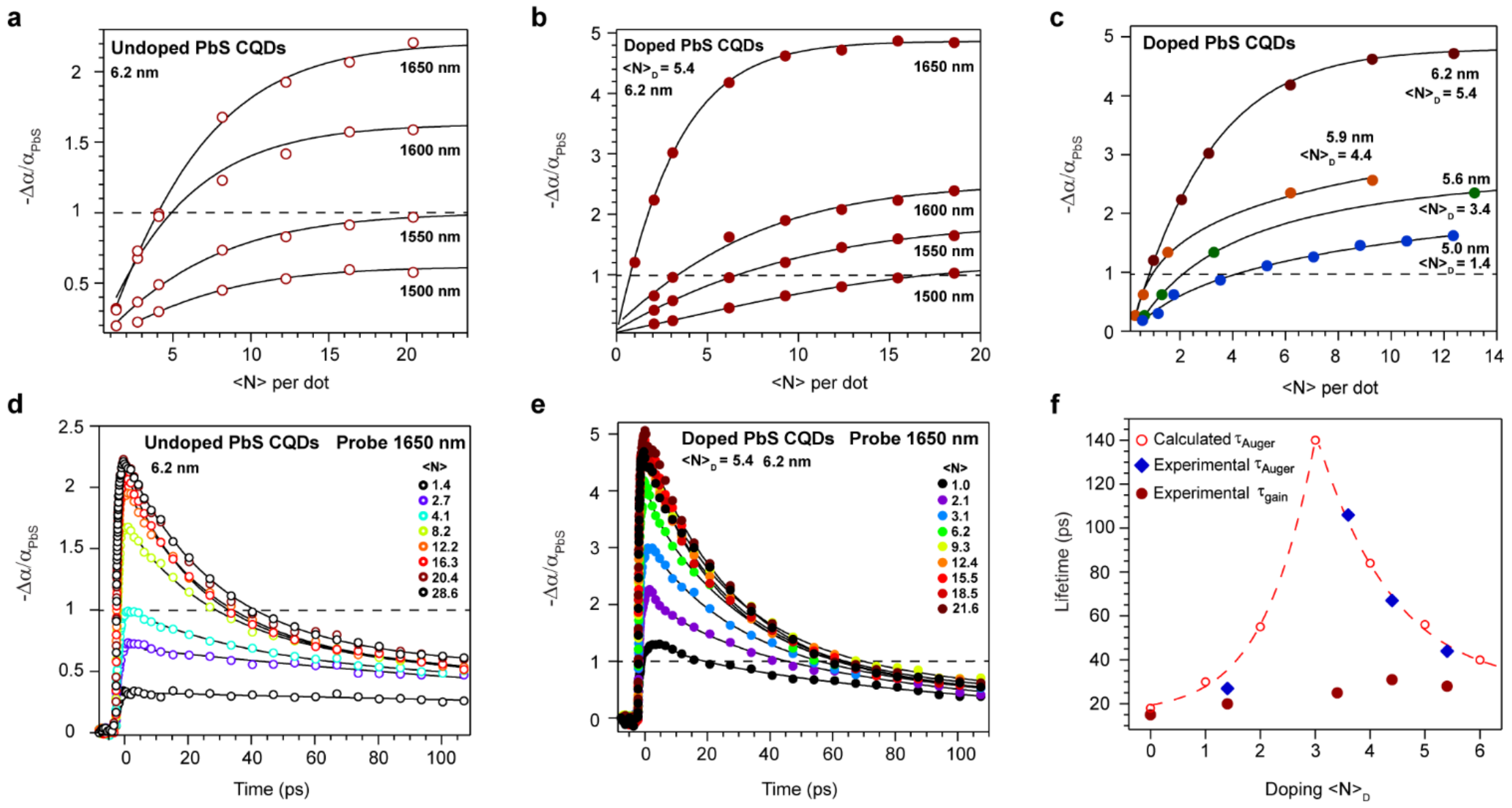

**Figure 2.** The gain spectra have been obtained in various probing wavelengths and different pump fluencies which are presented as exciton occupancy, <N> for both doped and undoped CQD films. The optical gain arises at the point that $-\Delta\alpha/\alpha_{PbS} > 1$ (horizontal black dash line) **a**, The gain spectra of the 6.2 nm PbS CQD (exciton peak at 1600nm) shows gain threshold at $\langle N \rangle_{thr} = 4$ **b**, The corresponding doped film exhibits a drastically reduced gain threshold at $\langle N \rangle_{thr} = 0.9$ at 1650 nm **c**, Comparison of gain spectra of various sized doped PbS CQDs films, possessing different initial doping values (the gain have been exctrated from TA data of Figure 2 d,e and supplementary section 10). The $\langle N \rangle_{thr}$ reduces upon increasing initial doping. **d**,**e** The transient absorption curves of the aforementioned doped and undoped samples at the probing wavelength of 1650 nm show that in the regime below optical gain the carrier dynamics are govern by Auger recombination and singe exciton (for the undoped samples) or charged single exciton (for the doped samples) relaxation. In the optical gain regime another fast component appears which is assigned to the gain lifetime **f**, Comparison of the calculated (open cycles) and the experimental (solid rhombus) Auger lifetime with the gain lifetime (solid cycles) at the gain threshold. The calculated Auger rates have been calculated using the experimental biexciton Auger lifetime of 210 ps ( measured for 5.5 nm PbS CQDs) at the exciton occupancy that reaches the gain threshold depending on the initial doping occupancy of the dots.

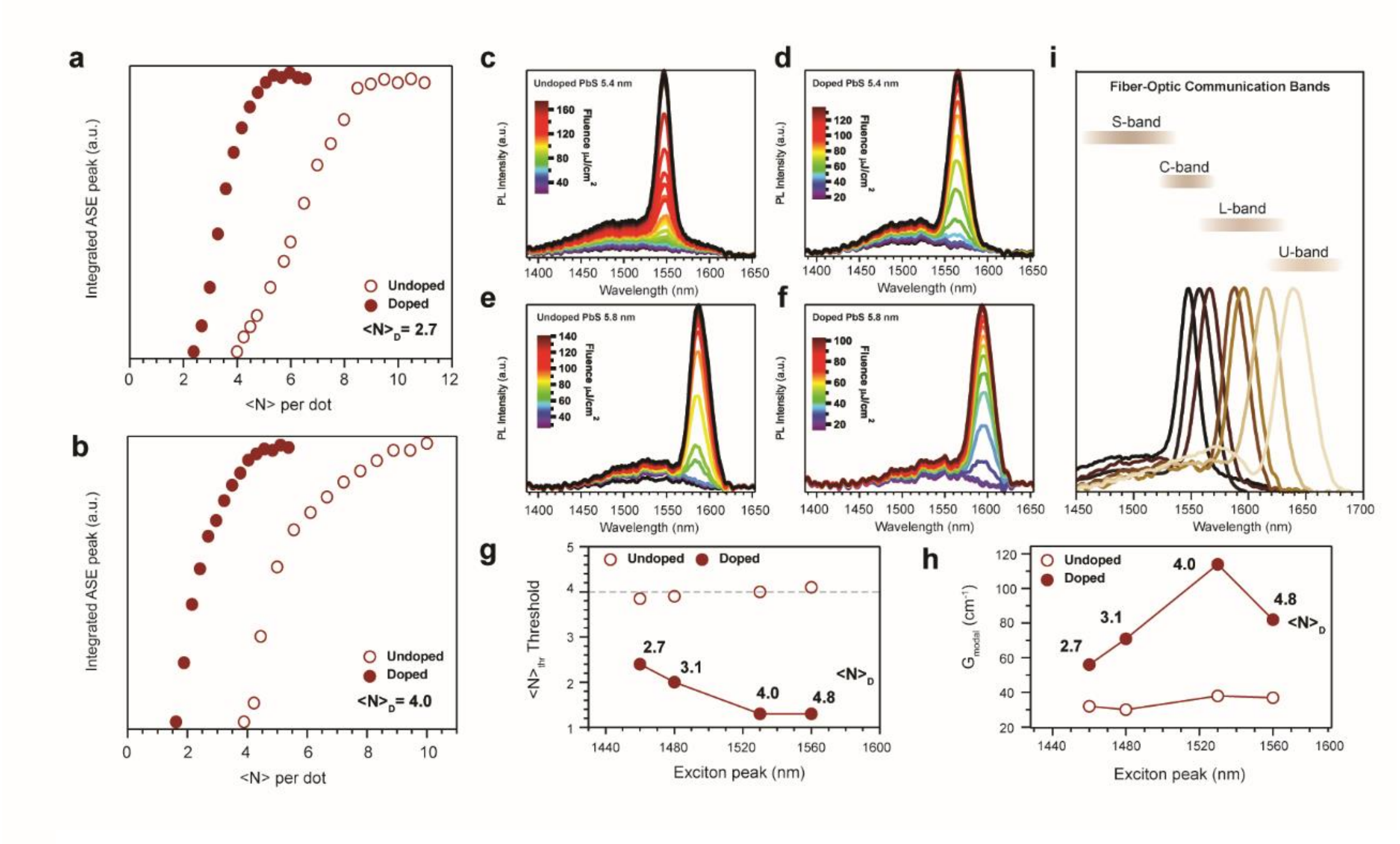

**Figure 3.** We collected the ASE and the VSL spectra in a series of PbS CQD films both doped and undoped, while here we show selective data for two samples with $\langle N\rangle_D$ of 2.7 and 4. We have excited our samples with a stripe-shaped laser beam at 800nm with a Ti:Saphire femtosecond laser with pulse width of 80fs and repetition rate of 1KHz at ambient conditions. **a**,**b**, The ASE area has been calculated by fitting the ASE peak with a Gaussian function while the photon density it is shown in terms of exciton occupancy. The solid red dot represent the integrated ASE area of doped and the hollow red dots the undoped samples. We further confirm that the undoped samples show ASE threshold $\langle N\rangle_{thr} = 4$, while the doped samples show $\langle N\rangle_{thr} < 4$ while both ASE signal are saturated for a total occupancy of 8 **c-f**, The corresponding PL spectra of these sample in a range of pumping intensities with a threshold of 70 μJ/cm$^2$ for the undoped and down to 25 μJ/cm$^2$ for the doped films **g**, Summarizing ASE thresholds in terms of exciton occupancy in PbS CQD of different sizes **h**, The net modal gain in the undoped films remains stable at 30 cm$^{-1}$ while in doped samples is increased with a peak value of 114 cm$^{-1}$ **i**, Collective ASE spectra from a series of PbS CQD films shows the tunability of the ASE peak from 1530 nm to 1650 nm, within the range of the fibre-optic communication bands.

TOC Figure

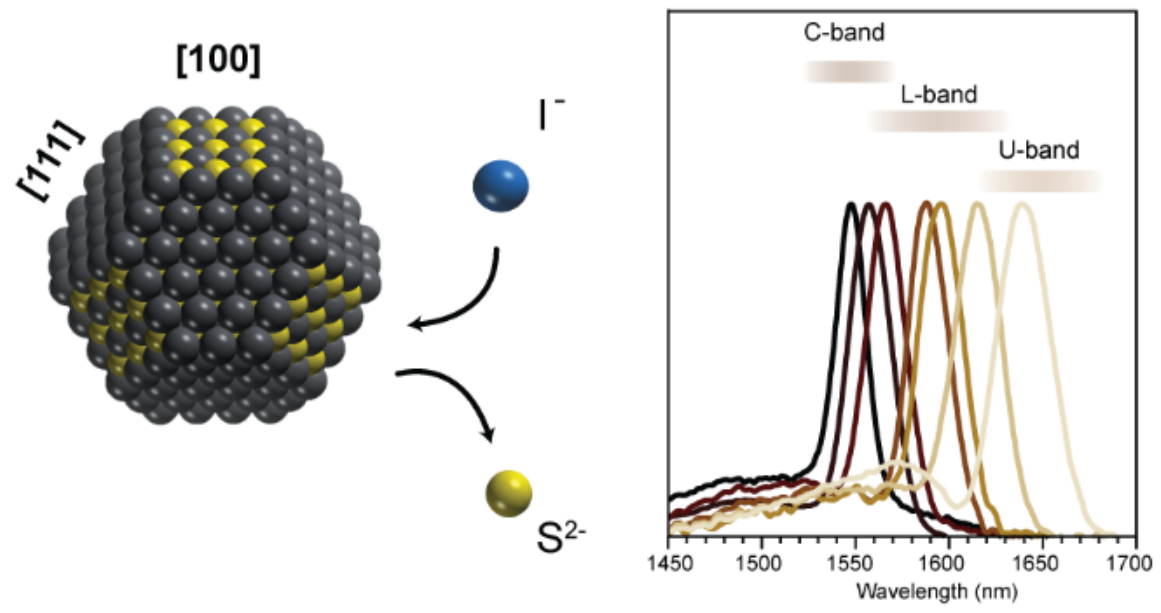